\UseRawInputEncoding
\documentclass[fleqn,usenatbib]{mnras}

\usepackage{newtxtext,newtxmath}

\usepackage[T1]{fontenc}

\usepackage{longtable}

\DeclareRobustCommand{\VAN}[3]{#2}
\let\VANthebibliography\thebibliography
\def\thebibliography{\DeclareRobustCommand{\VAN}[3]{##3}\VANthebibliography}

\usepackage{graphicx}	
\usepackage{amsmath}	

\newcommand{\feh}{\ensuremath{[\textrm{Fe}/\textrm{H}]}}
\newcommand{\kms}{\ensuremath{\textrm{km/s}}}
\newcommand{\fehcut}{FeH_{\rm cut}}

\newcommand{\dl}{Dulais Structure}

\def\red{\textcolor{black}}

\title[A recent accretion event in M31]{Chemo-dynamical substructure in the M31 inner halo globular clusters:\\Further evidence for a recent accretion event}

\author[G. F. Lewis et al.]{Geraint~F.~Lewis$^{1}$\thanks{E-mail: geraint.lewis@sydney.edu.au (GFL)},
Brendon~J.~Brewer$^{2}$,
Dougal~Mackey$^{3}$,
Annette~M.~N.~Ferguson$^{4}$, \newauthor 
Yuan~(Cher)~Li$^{2}$
\&
Tim~Adams$^{1}$
\\
$^{1}$Sydney Institute for Astronomy, School of Physics, A28, The University of Sydney, NSW 2006, Australia\\
$^{2}$Department of Statistics, The University of Auckland Private Bag 92019, Auckland 1142, New Zealand\\
$^{3}$Research School of Astronomy and Astrophysics, Australian National University, Canberra, ACT 2611, Australia \\
$^{4}$Institute for Astronomy, University of Edinburgh Royal Observatory, Blackford Hill, Edinburgh EH9 3HJ, UK
}

\date{Accepted XXX. Received YYY; in original form ZZZ}

\pubyear{2022}

\begin{document}
\label{firstpage}
\pagerange{\pageref{firstpage}--\pageref{lastpage}}
\maketitle

\begin{abstract}
Based upon a metallicity selection, we
identify a significant sub-population of the inner halo globular clusters in the Andromeda Galaxy which we name the \dl. 
It is distinguished as a co-rotating group of 10-20 globular clusters which appear to be kinematically distinct from, and on average more metal-poor than, the 
majority of the inner halo population.
Intriguingly, the orbital axis of this \dl\ is closely aligned with that of the younger accretion event recently 
identified using a sub-population of globular clusters in the outer halo of Andromeda, and this is strongly suggestive of a causal relationship between the two. 
If this connection is confirmed, a natural explanation for the kinematics of the globular clusters in the \dl\ is that they trace the accretion of a substantial progenitor ($\sim 10^{11} M_\odot$) into the halo of Andromeda during the last few billion years, that may have occurred as part of a larger group infall.
\end{abstract}




\section{Introduction}
Within our cosmological paradigm, large galaxies like the Milky Way have 
grown over time through the accretion of smaller stellar systems. The rate of this 
accretion is expected to be a relatively steady rain over cosmic time, potentially punctuated 
by more substantial accretion events~\citep{2002ApJ...568...52W}.
Since the discovery of the disrupting Sagittarius Dwarf Galaxy \citep{1994Natur.370..194I}, extensive surveys of
the halo of the Milky Way have revealed a wealth of substructure in the form of almost one hundred stellar streams \citep[e.g.,][]{2020ApJ...891L..19I,2022ApJ...928...30L,2022arXiv220410326M}, 
as well as more ancient extended structures \citep[e.g.,][]{2018ApJ...863L..28M,2018MNRAS.478..611B,2018Natur.563...85H,2020MNRAS.498.2472K}.
Furthermore, deep imaging of the Andromeda Galaxy (M31) has uncovered large-scale stellar substructure strewn throughout the halo, similarly revealing an extensive history of accretion
\citep[e.g.,][]{2018ApJ...868...55M}.
Recently, the  identification of kinematically distinct populations of globular clusters orbiting in the
outer halo of the Andromeda Galaxy, distinguished by their relationship to various underlying stellar
substructures, revealed two pronounced accretion episodes, one ancient and the other  more 
recent~\citep{2019Natur.574...69M}. 

The discovery of these distinct subgroups in the outer halo globular cluster population 
motivates us to revisit the kinematic properties of the inner globular clusters of Andromeda. 
Being the nearest large galaxy to our own, there has been significant interest in similarities and
differences between the Milky Way and Andromeda globular cluster populations \citep[e.g.][]{1969ApJS...19..145V,1982ApJ...259L..57H,1974ApJ...190..283H,1983IAUS..100..359F,1984ApJ...287..586B,1985ApJ...288..494C,1988ApJ...333..594E}, identifying a bimodel metallicity distribution \citep[e.g.][]{2000AJ....119..727B}, and potential small-scale kinematic and chemical substructures \citep[][]{2002AJ....123.2490P,2003ApJ...589..790P}.
Recently, \citet{caldwell:16} undertook an extensive spatial-kinematic-chemical analysis of the inner population of Andromeda's globular clusters based upon a homogeneous spectroscopic data set. Splitting their sample into a metal-rich, intermediate- and poor- sub-populations, they conclude that the kinematics of the metal-rich components are consistent with galactic rotation, whilst more metal-poor sub-components show a higher rotation and velocity dispersion and are more weakly concentrated with regards to galactic disk, and hence appear to represent a transition to the larger halo population of globular clusters. However, we note that the kinematic analysis of \citet{caldwell:16} was undertaken with respect to rotation about the minor axis of Andromeda, without exploration of the potential orientation of the rotation.

Hence, here we undertake a more expansive exploration of the kinematic and chemical distributions of the inner globular cluster populations in Andromeda, including additional freedom to explore the axes of rotation. 
As detailed below, the results of this analysis is the identification of a pronounced substructure within the inner globular cluster 
population, which we name the
\dl\footnote{Pronounced `dill'+`ice', after the Welsh river name meaning `black stream'.}; intriguingly
the kinematic and spatial properties causally link it to the recent accretion events in the outer halo 
identified by \citet{2019Natur.574...69M}.
The layout of this paper is as follows: In Section~\ref{sec:approach} we outline our approach, including the data employed and the statistical methods used in model assessment. We discuss the results of this study in ~\ref{sec:discussion} and present our conclusions in Section~\ref{sec:conclusions}. 

\section{Approach}
\label{sec:approach}
In our previous exploration of the kinematic properties of the globular clusters in the outer halo of M31 (i.e., at projected galactocentric radii $> 25$\ kpc), we defined two distinct populations according to whether or not the clusters were associated with underlying stellar substructure \citep{2019MNRAS.484.1756M}. 
This indicated whether the globular clusters were accreted relatively recently, such that their progenitors are still undergoing tidal disruption, or whether they potentially trace more ancient accretions which have been completely dispersed. However, it is impossible to identify faint stellar substructures against the stellar disk and extensive tidal debris within the more central regions of M31 \citep{2018ApJ...868...55M}, meaning that this experiment cannot easily be repeated for the inner halo population. Instead, our present study focuses upon splitting the inner clusters according to their chemical enrichment, or metallicity, in order to assess whether there are kinematically distinct sub-groups inhabiting the inner regions of Andromeda.

We adopt a Bayesian modelling approach, providing 
 an assessment of Bayesian evidence (marginal likelihood) and allowing a determination of the relative efficacy of various models in fitting the data. The goal of our analysis is to compare whether a single component model, or a dual component model separating the globular clusters based upon their metallicity, provides a more plausible explanation of the data.

As detailed below, we assume that the kinematics of a given globular cluster population can be represented by a rotational component coupled with a velocity dispersion, and consider three distinct forms for the rotation. 
In the first stage of the analysis, we fit models to the total globular cluster population, calculating the Bayesian evidence for each rotational form. We then repeated the process to explore whether two distinct populations, split by metallicity and each with their own rotational component and velocity dispersion, represented a better fit of the data.
Given the relatively large uncertainties on the available metallicity measurements, we adopted a probabilistic approach to the two-population "split", weighting each globular cluster's contribution to a kinematic component according to its metallicity and associated uncertainty relative to a metallicity threshold (treated as a free parameter).  For the sake of clarity, the following discussion refers to globular clusters with metallicites lower than the metallicity cut as metal-poor, whilst those above are referred to as metal-rich.

\subsection{Globular Cluster Data}
\label{subsec:data}
The data underpinning our analysis is the publicly-available catalogue of \citet{caldwell:16}, originally obtained through spectroscopic observations for nearly all of the known globular clusters in the inner regions of Andromeda using the Hectospec multi-fibre spectrograph \citep{fabricant:05} on the 6.5m MMT Observatory telescope in Arizona. 
\citet{caldwell:16} used these spectra to provide a set of uniformly-derived line-of-sight velocities and metallicity estimates for approximately $94\%$\ of globular clusters with projected galactocentric radii inside $21$\ kpc. Our previous analysis of globular clusters in Andromeda's outer halo \citep{2019MNRAS.484.1756M,2019Natur.574...69M} had an inner cut-off of $25$\ kpc, so we searched the literature, and in particular the cluster catalogue from the Pan-Andromeda Archaeological Survey \citep[PAndAS;][]{2014MNRAS.442.2929V}, to ensure we were not missing any objects in the range $21-25$\ kpc. We also updated the velocities from \citet{caldwell:16} with more precise PAndAS measurements for a few clusters (G268, H14, and SK255B). In summary, the total base catalogue comprises 345 globular clusters extending to a projected galactocentric radius of $25$\ kpc.

\citet{caldwell:16} convincingly established that the most metal-rich globular clusters in the central regions of M31 are kinematically and spatially associated with its disk. Since our analysis is focused on Andromeda's inner halo, we removed this component by excluding all clusters with metallicity $FeH \geq -0.4$\footnote{Throughout we use $FeH$ to represent \feh.}.
We further removed all globular clusters for which the observations were of insufficient quality to determine a reliable metallicity estimate. 

\begin{table}
\centering
\begin{tabular}{ll}
\hline
Models  $(\rm{I}\ \&\ \rm{II})$   & Functional form \\
\hline
$V$     & $A_k\sin(\theta - \phi_k)$ \\
$S$    & $A_k\left(x\sin\phi_k - y\cos\phi_k\right)$ \\
$F $    & $A_k\tanh\left((x\sin\phi_k - y\cos\phi_k)/L_k\right)$ \\
\hline
$\top$     & Disjunction of all of the above $(\rm{I}\ \&\ \rm{II})$ \\
\hline
\end{tabular}
\caption{The three different rotational model components considered in our kinematic models, providing the line-of-sight velocity as a function of position on the sky.
The positions are expressed in either Cartesian coordinates $(x,y)$ or plane polar
coordinates $(r,\theta)$ defined as specified in the text.
A single roman numeral subscript (e.g., $V_{\rm I}$) corresponds 
to a single-population model whose parameters are represented 
with a single subscript $k=0$ (e.g., $A_0$). 
For models with two populations, (e.g., $V_{\rm II}$), the globular cluster subgroup below the metallicity cut is represented by subscript $k=0$, whereas the globular cluster subgroup above the metallicity cut is represented by a subscript $k=1$. The model $\top$ is the proposition that one of the three rotational forms is correct.
\label{tab:models}}
\end{table}

The final sample for consideration contains $n=278$ globular clusters. Each cluster possesses the following data: $(x_i,y_i,v_i,\sigma_{v_i},\widehat{FeH_i},\sigma_{FeH_i})$ -- namely the Cartesian tangent-plane position centred on Andromeda, $(x_i,y_i)$, 
the Andromeda-centric line-of-sight velocity and associated uncertainty, $(v_i,\sigma_{v_i})$, and the reported metallicity and associated uncertainty, $(\widehat{FeH_i},\sigma_{FeH_i})$; 
the metallicity and velocities of the globular cluster population, coupled with their uncertainties, is presented graphically in the Appendix as Figure~\ref{fig:comparison}.
Cartesian positions correspond to tangent-plane distances centred on M31; these coordinates, and the Andromeda-centric velocities, were calculated according to the prescription outlined in \citet{2019MNRAS.484.1756M}. Note that the transformation of Cartesian coordinates into regular polar coordinates is defined as:
\begin{equation}
    r_i = \sqrt{x_i^2 + y_i^2} \ \ \ \ \ \ \ \ \ \ \ \ \theta_i = \arctan{2( y_i , x_i )}.
    \label{eqn:geometry}
\end{equation}
Here, $\theta_i$ is not the position angle $(PA)$ on the sky that is usually quoted in astronomy, which is measured from North through East, but instead is defined from East through North. %
For convenience, the results of this study are converted into astronomical position angles 
at the completion of the analysis.

The typical individual $1\sigma$ uncertainty on $\widehat{FeH_i}$ is $\sigma_{FeH_i}=0.2$\ dex, while that on the Andromeda-centric velocity $v_i$ is $\sigma_{v_i}=7.5$\ km/s. This latter value is dominated by the measurement uncertainty (typically $\approx6$\ km/s), with a small contribution due to the assumed systemic motion of M31 \citep{2019MNRAS.484.1756M}.

\begin{table}
\centering
\begin{tabular}{l|l}
Parameter & Prior \\
\hline
$\fehcut$     & Uniform(-3, -1) \\
$A_k$           & Uniform(0, 800) km/s \\
$\phi_k$        & Uniform$(-\pi, \pi)$ radians \\
$L_l$           & Uniform$(0, 2)$ degrees \\
$\Sigma_{k0}$ & Uniform(0, 400) km/s \\
$s_k$         & Uniform(-4, 4) (degrees)$^{-1}$
\end{tabular}
\caption{Prior distributions for the unknown parameters in our models.
The $L$ parameter only applies to the model with asymptotically flat rotation ($F$).\label{tab:priors}}
\end{table}

\subsection{Kinematic Modelling}
\label{subsec:dynamicalmodelling}
As with previous studies \citep[c.f.][]{2019Natur.574...69M}, we explored three different kinematic models.
Each of these is a nonlinear regression model that predicts line-of-sight
velocity $v_i$ as a function of position $(x_i, y_i)$ or $(r_i, \theta_i)$, and potentially (when there are two populations) the metallicity $FeH_i$. The three models are denoted $V,S,F$ and possess different forms for the rotational component as presented in Table~\ref{tab:models}; note that these models will be subscripted with ${\rm I}$ and ${\rm II}$ if they represent models with a single or double kinematic components.
$V$ is a simple sinusoidal function previously used in a variety of studies of the rotation of the globular cluster population in the halo of Andromeda \citep[e.g.,][]{2013ApJ...768L..33V,2014MNRAS.442.2929V,caldwell:16}, 
whereas $S$ corresponds to a solid-body rotation. Finally, $F$, represents an asymptotically-flat 
rotation curve which smoothly transitions from one side of the galaxy to the other.
For a single-population model (e.g. $V_{\rm I}$), the index $k$ is fixed at $0$. When the metallicity cut is applied and the model as two kinematic components (e.g. $V_{\rm II}$), the globular cluster population below the metallicity cut is represented by subscript $0$, whereas the globular cluster population above the cut is represented by subscript $1$.

As well as rotation, it is also assumed that each globular cluster population possesses a velocity dispersion. We adopt the functional form of the velocity dispersion over the populations to be
\begin{equation}
    \Sigma_k( r ) = \Sigma_{k0} \exp{ \left( s_k r \right) }
    \label{eqn:velcoitydispersion}
\end{equation}
where $\Sigma_{k0}$ and $s_k$ are parameters to be determined; clearly if $s_k = 0$ then the velocity dispersion
is spatially constant over the population.

\subsection{Metallicity Selection}
\label{subsec:metallicityselection}
The models considering two populations assume that there exists a metallicity threshold
$\fehcut$ that cleaves the globular cluster sample into two dependent subgroups based upon whether an individual globular cluster's metallicity, ${FeH_i}$, 
is above or below this limit. However, as the metallicity uncertainties, $\sigma_{FeH_i}$, can be significant, this is analogous to a curve fitting
problem with error bars on the $x$-values  \citep{hogg_straightline}.
Thus, we included all $n=278$ true metallicities as extra parameters to be inferred.
We take the prior for each true metallicity to be a normal distribution centered at the measured value and with the appropriate standard deviation:
\begin{align}
FeH_i &\sim \textnormal{Normal}\left(\widehat{FeH_i}, \sigma_{FeH_i}^2 \right),
\end{align}
where $\widehat{FeH_i}$ is the given measurement and $\sigma_{FeH_i}$
the reported uncertainty.

For simplicity we did not use a hierarchical
model, so the metallicities do not 'borrow strength' from each other (i.e.,
there is no tendency for an unknown metallicity value to be dragged towards the
preponderance of the other metallicities).

\begin{figure*}
    \centering
    \includegraphics[width=0.95\textwidth]{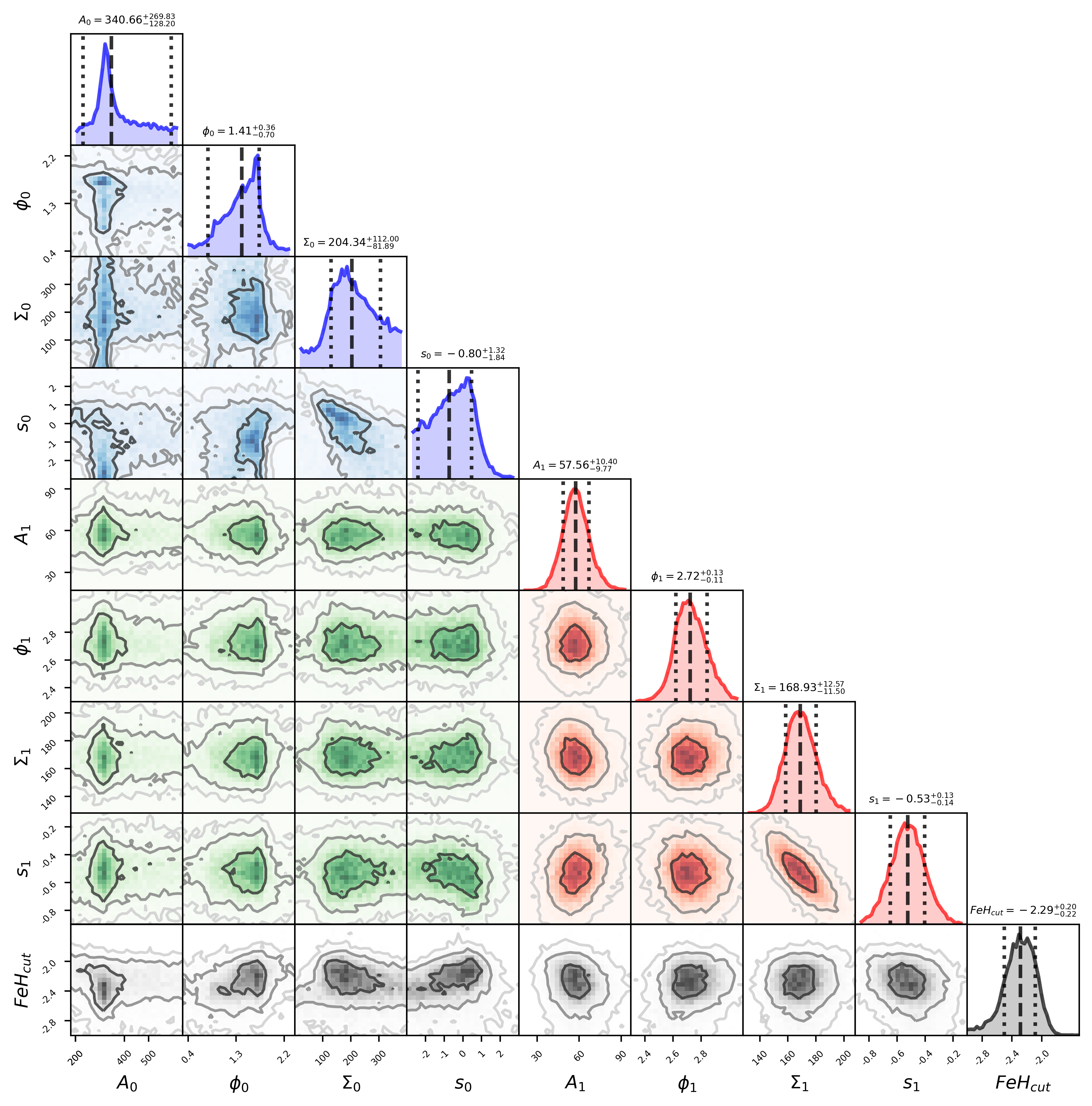}
    \caption{Corner plot for the best-fit model as outlined in the text. Here, the blue distributions correspond to the metal-poor component of the globular cluster population, whilst the red represents the metal-rich component. The green distributions show the relationship of the parameters of the two populations against each other. The final row presents the metallicity cut. Note that the $F_{\rm II}$ kinematic model has two additional scale-length parameters $L_k$, which are constrained to be relatively small and are omitted from the plot for clarity. The best-fit model parameters are tabulated in Table~\ref{tab:margin}.
    }
    \label{fig:corner}
\end{figure*}

\subsection{Likelihood Calculation}
\label{subsubsec:likelihoodcalculation}
For a given rotational model $M_j = \{ V_j,\,S_k,\,{\rm or}\ F_j \}$ (where $j = {\rm I\ or\ II}$) which are dependent upon a set of parameters $\Theta_k$, 
and assuming a normal distribution for the velocity uncertainties on the globular clusters, we define the corresponding likelihood to be
\begin{equation}
    {\cal L}_k( {\cal D}_i) = \frac{1}{\sqrt{2\pi }\sigma_{ki}} 
    \exp{\left( -\frac{(M_j({\cal D}_i|\Theta_k) - v_i)^2}{2\sigma^2_{ki}}\right) }\,,
    \label{eqn:modellikelihood}
\end{equation}
where the velocity dispersion and the measurement uncertainty are added in quadrature such that 
\begin{equation}
    \sigma_{ki}^2 = \sigma_{v_i}^2 + \Sigma_k^2( {\cal D}_i|\Theta_k )\,.
\end{equation}
When considering a single-population model, such that $k=0$, the total likelihood is given by
\begin{equation}
    {\cal L} = \prod_{i=1}^n  {\cal L}_{0}({\cal D}_i|\Theta_0)\,.
    \label{eqn:likelihood}
\end{equation}
However, when considering a model with two populations split by metallicity such that $k=0,1$,
it is appropriate to select the corresponding kinematic component for the likelihood term in the product.
\begin{equation}
    {\cal L} = \prod_{i=1}^n \left\{
        \begin{array}{lr}
    {\cal L}_{0}({\cal D}_i|\Theta_0), & FeH_i < \fehcut \\
    {\cal L}_{1}({\cal D}_i|\Theta_1), & FeH_i \geq \fehcut \\
            \end{array}
    \right.
    \label{eqn:likelihoodprod}
\end{equation}
As typical in numerical calculations involving likelihood,
practical considerations require that the log of the likelihood is computed.

For each model, we determine the posterior distribution for the parameters
given the data, and the marginal likelihood of the model as a whole,
using Diffusive Nested Sampling  \citep{dns, dnest4}, a variant of
Nested Sampling  \citep{2004AIPC..735..395S,10.1214/06-BA127}.

\subsection{Priors}
\label{subsec:influencepriors}
For simplicity, we employed uniform priors for all parameters, with astronomically informed limits; these are given in Table~\ref{tab:priors}.
In Bayesian model comparison, the priors can have a significant effect on the marginal likelihoods.
As a consequence, the marginal likelihoods we compute in this study are sensitive to the prior widths. 
However, this is not a major concern for two reasons.
Firstly, we used the same priors for all parameters shared in common between models so they have similar predictive ability, and
secondly, the main caveat to Bayesian model comparison with uniform priors is that very
wide limits might overly penalise the more complex model.
Since we found that the two-component model was favoured despite the wide uniform priors,
a more sophisticated choice of priors would likely strengthen this result, rather than weakening it.

\begin{table}
\centering
\begin{tabular}{lll}
Model & $\ln Z$ & $Z/Z_{\rm max}$\\
\hline
$V_{\rm I}$  & -1780.76 & 0.051 \\
$S_{\rm I}$ & -1785.82 & 0.00032  \\
$F_{\rm I}$  & -1780.27 & 0.083 \\
\hline
$V_{\rm II}$ & -1778.39 & 0.54 \\
$S_{\rm II}$ & -1784.79 & 0.00090 \\
$F_{\rm II}$ & {\bf-1777.78} & 1\\
\hline
$\top$ & -1779.05
\end{tabular}
\caption{Marginal likelihoods or evidences for the three single-component models followed by the three
two-component models. The flat kinematics $F$ predicts
the data best, followed closely by the $V$ form. The solid-body $S$ kinematic model is heavily
disfavoured. Two components are favoured over a single component model in every case, by
Bayes factors of around $\exp(2) \approx 10$.
Finally, we present the marginal likelihood for the top statement
$\top = V_{\rm I} \vee S_{\rm I} \vee F_{\rm I} \vee V_{\rm II} \vee S_{\rm II} \vee F_{\rm II}$,
which is the logical disjunction of all six models, assuming $1/6$ prior probability each. This is the evidence for the statement that one of these six models is true.\label{tab:logzs}}
\end{table}

\section{Results and Discussion}
\label{sec:discussion}

\begin{figure*}
    \centering
    \includegraphics[width=0.99\linewidth]{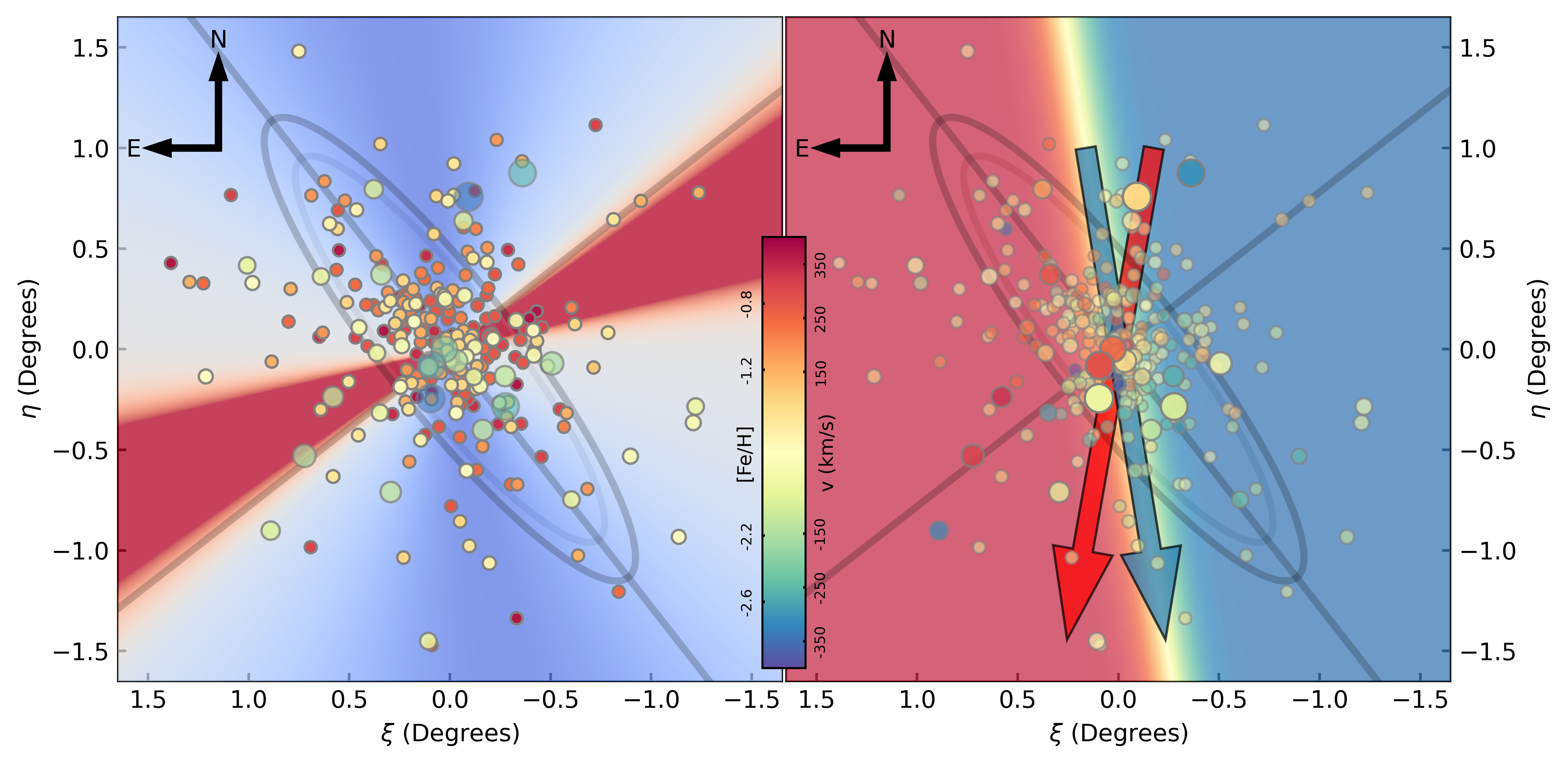}
    \caption{The best-fit model for the inner globular cluster population of the Andromeda Galaxy (c.f. Figure~\ref{fig:corner}). In each panel, the globular clusters are represented as filled circles colour-coded with their metallicity in the left-hand panel, and with velocity in the right-hand panel. The size of each circle reflects the metallicity of each globular cluster, with more metal-poor clusters being larger (scaling is set by a logistic function centred at $FeH=-2.2$ and a width of 0.2 in metallicity). The transparency of each globular cluster symbol in each panel has been adjusted based upon metallicity, such that the left-hand panel emphasises the metal-rich component, whereas the right-hand panel emphasises the metal-poor population.
    The disk of the Andromeda Galaxy is represented as ellipses. In the left-hand panel, the orientation of the two rotational components are presented, with the metal-poor axis of rotation in blue and the metal-richer component in red, with the width of the distributions representing the uncertainty on these quantities from the Bayesian analysis. In the right-hand panel, the kinematic model is represented, with the blue arrow representing the angular momentum of the population. The red arrow is the angular momentum of the outer halo globular cluster population that is associated with substructure and represents a recent accretion into Andromeda.  
    Note that, for the purposes of this figure, the tangent plane coordinates are labelled with $(\xi,\eta)$ to aid comparison to previous work. However, for the analysis presented in this paper, $(x,y)$ are used to represent these coordinates.
    }
    \label{fig:model}
\end{figure*}

\subsection{Bayesian Evidence}
\label{subsec:bayesian}
The Bayesian evidences from our nested sampling exploration of the posterior probability spaces are presented in Table~\ref{tab:logzs}. One immediate result is that, irrespective of the rotational model under consideration, the Bayesian evidence explicitly favours the two component rotational model over the single component, with a separation of the populations at metallicity $FeH \sim -2.2$ consistent across the models. For the best-fitting model, $\{F_{\rm II} \}$, the ratio of the Bayesian evidence between the single and two components fits, known as the Bayes factor, is 12.06, with it taken as "strong" evidence that the two component model is the better description of the data \citep{10.2307/2291091}.
We summarise the properties of the best-fit model in the corner plot presented 
in Figure~\ref{fig:corner} and tabulated in Table~\ref{tab:margin}; again, we emphasise that the characteristics of this best-fit are seen across all of the two component fits. Essentially, the 
 the metal-poor component, which consists of $\sim 10-20$ globular clusters, has a rotational velocity amplitude of $\sim340 \kms$, substantially higher than the more metal-rich population (with a rotational amplitude of $\sim55 \kms$),
 \red{although we note that the large rotational amplitude is potentially driven by a small number of globular clusters at the velocity extremes.}
 Furthermore, the velocity dispersion of the metal-poor component, at $\sim200 \kms$, is  higher than that for the more metal-rich component $(\sim170 \kms)$. Whilst these values refer to the $50^{\rm th}$ percentile of the posterior probability distributions, it should be noted that some of these show substantial skewness, and we note that with this the velocity dispersions are roughly consistent within the uncertainties. \red{We comment on this potentially large velocity dispersion in the conclusions to this paper.}

\begin{figure*}
    \centering
    \includegraphics[width=0.80\linewidth]{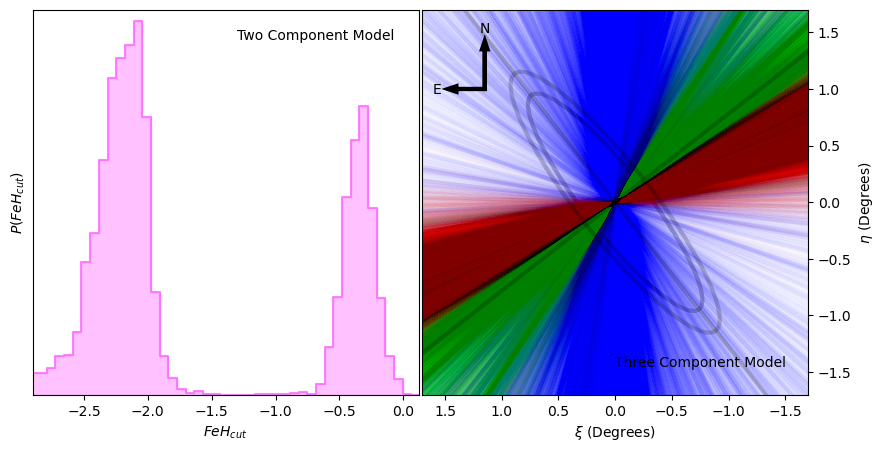}
    \caption{\red{Summary of the analysis of the entire globular cluster population, including those with $FeH \ge -0.4$ (see Section~\ref{subsec:data}). The left-hand panel presents a two component model  where the prior range of the metallicity cut was expanded to encompass the entire span of metallicity, showing distinct peaks at $FeH \sim -2$ and $FeH \sim -0.4$. The right-hand panel presents the distribution of rotational axes from a three component model with hard metallicity cuts at $FeH = -2.0$ and $FeH = -0.4$,  with $FeH < -2.0$ in blue, $-2.0 \le FeH \le -0.4 $ in red and $FeH > -0.4$ in green. Clearly, the rotational axes of the sample with $FeH > -2.0$ are aligned with the rotational axis of the Andromeda Galaxy, whereas the more metal-poor sample possesses a distinctly offset rotational axis; it is this we identify as the \dl.}
    }
    \label{fig:bayes}
\end{figure*}

\begin{table}
\centering
\renewcommand{\arraystretch}{1.2} 
\begin{tabular}{l|rlc}
Parameter & \multicolumn{2}{c}{Value} & Units \\
\hline
$A_0$ & $340.66$ & $^{+269.83}_{-128.20}$ & km/s\\
$\phi_0$ & $ 1.41$ & $^{+ 0.36}_{- 0.70}$ & radians\\
$\Sigma_0$ & $204.34$ & $^{+112.00}_{-81.89}$ & km/s \\
$s_0$ & $-0.80$ & $^{+ 1.32}_{- 1.84}$ & (degrees)$^{-1}$\\
$L_0$ & $ 0.77$ & $^{+ 0.79}_{- 0.62}$ & degrees\\
$A_1$ & $57.56$ & $^{+10.40}_{- 9.77}$ & km/s\\
$\phi_1$ & $ 2.72$ & $^{+ 0.13}_{- 0.11}$ & radians\\
$\Sigma_1$ & $168.93$ & $^{+12.57}_{-11.50}$ & km/s\\
$s_1$ & $-0.53$ & $^{+ 0.13}_{- 0.14}$ &  (degrees)$^{-1}$\\
$L_1$ & $ 0.13$ & $^{+ 0.09}_{- 0.07}$ & degrees\\
$FeH_{cut}$ & $-2.29$ & $^{+ 0.20}_{- 0.22}$\\
\end{tabular}
\caption{
The best-fit parameter values for the favoured $F_{\rm II}$ model, shown graphically in the corner plot in Figure~\ref{fig:corner}. The values
reflect the $50^{th}$ quantile of the posterior distribution, with the uncertainty determined from the $16^{th}$ and $84^{th}$ quantiles.
\label{tab:margin}}
\end{table}

\red{Before closing this section, we return to the fact that metallicity cut was made to the initial sample at $FeH \ge -0.4$ to remove the galactic component
identified by \citet[][see Section~\ref{subsec:data}]{caldwell:16}. To ensure this selection has not biased the findings in this paper, several additional tests were undertaken which are summarised here. Essentially, single and multiple rotational components, using model $F$ (see Table~\ref{tab:models}), were fit for each of the components, and the Bayesian evidence calculated. In fitting a single rotational component, the resultant Bayesian evidence $Z_1 = −2165.62$. The second model comprised of two components, but the prior on the metallicity cut was expanded to cover the entire metallicity range. The resultant evidence for this was $Z_2 = −2162.58$, demonstrating that this is favoured over the single component model. However, the posterior for the metallicity cut, presented in the left-hand panel of Figure~\ref{fig:bayes}, possesses two distinct peaks, one at $FeH \sim -2$ and the other at $FeH \sim -0.4$, indicating that the modelling suggests that the globular cluster populations could be split into three. Fixing the metallicity cut at $FeH = -2.0$ recovers the results presented in this paper, namely that there are two kinematically distinct populations with significantly different rotational axes. However, fixing the metallicity cut at $FeH = -0.4$ shows that both populations share a common axis of rotation, with those at $ FeH > -0.4 $ possessing a rotation amplitude of $\sim 55$ km/s, whilst the lower metallicity population possesses a rotational amplitude of $\sim 160$ km/s. Finally, a three component model was fit to the entire population, with fixed metallicity cuts at $FeH = -2.0$ and $FeH = -0.4$. The resulting Bayesian evidence was $Z_3 = −2156.5$, demonstrating that this is the favoured model. In the right-hand panel, we present the rotational axes of the three components from the posterior exploration, with $FeH < -2.0$ in blue, $-2.0 \le FeH \le -0.4 $ in red and $FeH > -0.4$ in green. The distribution of red and green axes overlap and are aligned with rotational axis of Andromeda, whilst the most metal-poor component in blue is distinctly offset from these. We conclude that disregarding globular clusters with $FeH > -0.4$ has not biased the findings in this paper.
}

\subsection{Exploring the Two Component Model}
\label{subsec:twocomponentmodel}
If we assume that the inner globular clusters of Andromeda represent a mix of those formed in-situ coupled with phase-mixed accretion events, a natural explanation for the properties of the metal-poor globular clusters is that it is simply a kinematically hotter component of the overall population. However, if this was the correct interpretation, we would expect the orientation of the rotation of the globular cluster population to reflect that of the overall Andromeda Galaxy. To explore this, we present the orbital orientation of the two components for the best-fit model is displayed in Figure~\ref{fig:model}. The globular clusters are shown shown as filled circles colour-coded with their metallicity (left-hand panel) and velocity (right-hand panel). In both panels, more metal-poor clusters being larger. In the left-hand panel, the more metal-rich globular clusters are emphasised through their opacity, whereas it is the more metal-poor clusters emphasised in the right-hand panel.
Underlying the globular clusters are solid lines denoting the major and minor axes of the Andromeda Galaxy, emphasised with overlying ellipses.

In the left-hand panel of Figure~\ref{fig:model} the axes of rotation of the metal-poor component (blue) and metal-rich component (red) are presented underlying the globular cluster population; again these are taken from the $50^{\rm th}$ percentile of the posterior probability distribution, with  one $\sigma$ uncertainties corresponding to the $16^{\rm th}$ and $84^{\rm th}$ percentiles (see Figure~\ref{fig:corner}). The rotational axis of the metal-rich component is reasonably aligned with the minor axis of the Andromeda Galaxy, and so we conclude that their kinematic properties are related to the larger-scale galactic rotation.     
However, the rotational axis of the  metal-poor component is substantially offset from the minor axis of the Andromeda Galaxy by $\sim75\degr$. This is emphasised in the angular momentum vectors of the two components, with the metal-rich population  at $PA\sim115\degr$ whereas the metal-poor component is at $PA\sim190\degr$\footnote{
The best-fit values for the orientation of the rotation 
are $\phi_0=1.41$ and $\phi_1=2.72$ radians for the \dl\ and 
dominant galactic population respectively (Table~\ref{tab:margin}). However, given the form of the 
underlying model (Table~\ref{tab:models}), the direction of
the rotational vector (given the right-hand rule) is these 
values $\pm \pi$. This corresponds to $\theta_0 \sim -100\degr$ 
and $\theta_1 \sim -25\degr$ in the model coordinates (Equation~\ref{eqn:geometry}). Given that $\theta$ and $PA$ 
are related through $\theta + PA = 90\degr$, this gives the quoted
$PA$s for the orientations of the spin vectors of the two 
components.
}
Hence, we concluded that the metal-poor globular clusters reveal the presence of a kinematically distinct population from the overall inner globular clusters; it is this that we identify as the  
\dl.
 We do, however, add a word of caution in interpreting these findings on the metallicity of the \dl\ as it is unlikely that globular cluster populations can cleanly be separated by metallicity alone and some of the low metallicity clusters are likely to be drawn from the galactic population, whereas the \dl\  will have globular clusters above $\fehcut$. What the results of this study show is that the kinematics of the \dl\  dominate at low metallicity, whereas it is lost against the kinematics of the galactic population at higher metallicities. This does indicate that the \dl\  is, overall, more metal-poor than the galactic population of globular clusters. We explore this in more detail below.

\begin{figure}
    \centering
    \includegraphics[width=0.95\linewidth]{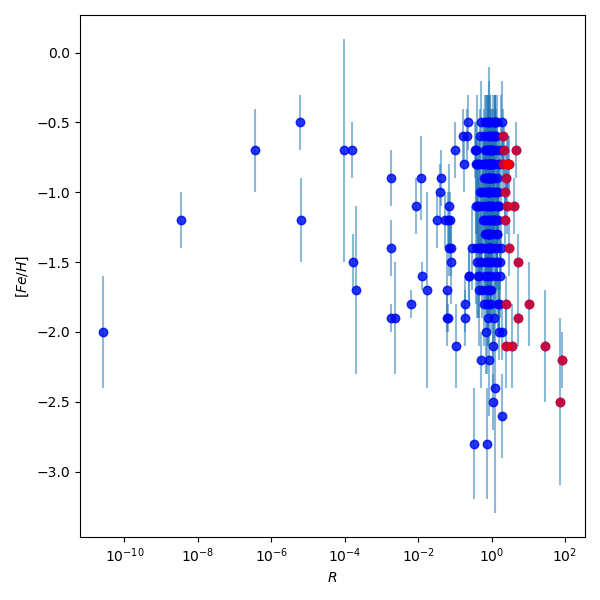}
    \caption{The ratio of the likelihoods, $R$, drawn from the two kinematic components of the best-fit model for the globular clusters in our sample, presented as a function of metallicity. The smaller the value of $R$, the more likely the globular cluster is drawn from the galactic populations. Those with $R<2$ are coloured blue, whereas those with $R>2$ are red.
    }
    \label{fig:rsplit}
\end{figure} 

 The right-hand panel of Figure~\ref{fig:model} presents the best-fit rotational model for the \dl\ component, colour-coded for velocity; clearly the direction of rotation is distinctly different to the rotation of Andromeda, which is aligned along the minor axis. Also in this figure, the blue arrow denotes the direction of the angular momentum vector for this rotation of the \dl\ component. The red arrow, however, presents the direction of the angular momentum for the outer globular cluster population that was associated with underlying stellar substructure, and therefore indicative of a relatively recent accretion event into the halo of the Andromeda Galaxy \citep{2019Natur.574...69M}. Clearly, given that the direction and rotational sense of these inner and outer globular cluster populations are extremely similar, it is natural to suggest that the two components are related. One obvious conclusion is that the \dl\  represents the remnants of a stellar system accreted as part of the larger-scale accretion seen in the outer halo of Andromeda.   

\begin{figure}
    \centering
    \includegraphics[width=0.95\linewidth]{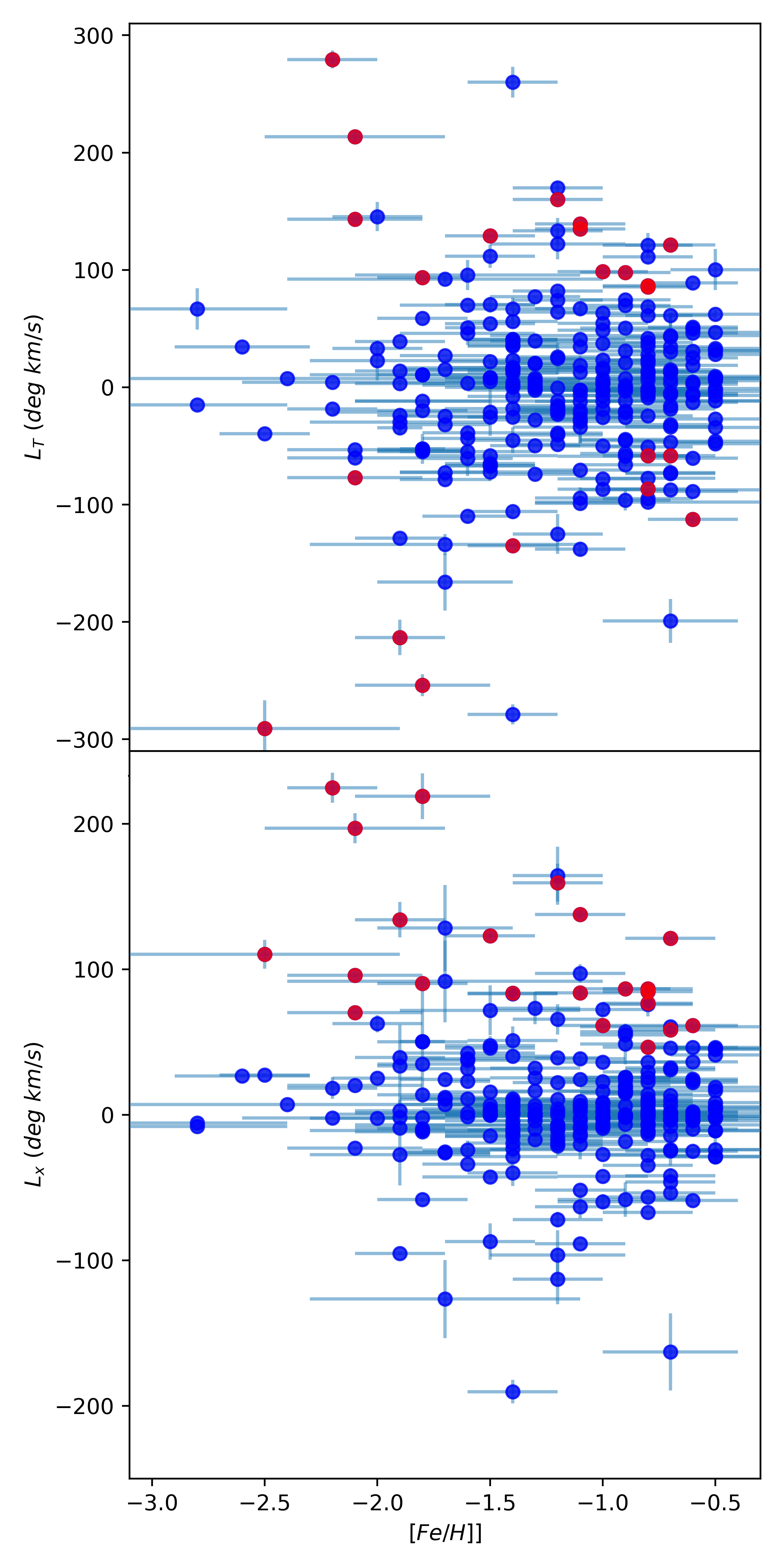}
    \caption{The total angular momentum, $L_T$, (top panel) and x-component of the angular momentum, $L_x$ (bottom panel) of the globular cluster population, colour-coded as described in Figure~\ref{fig:rsplit}. The upper panel reveals that those globular clusters more likely to be members of the \dl\ (coloured red) are at larger absolute angular momentum than the overall population. The lower panel illustrates that those globular clusters more likely members of the \dl\ are clustered at positive values and are distinct from the overall galactic population. We note that the significant clustering in $L_x$ is apparent as the rotational axis of the \dl\ is aligned with the y-axis of the tangent plane coordinates. 
    }
    \label{fig:pair}
\end{figure}

\subsection{Posterior Analysis}
\label{sub:posterior}
As discussed previously, given the limitations of the data the sub-populations of globular clusters were defined solely on a metallicity cut. However, care must be taken in interpreting the metallicity distribution of the \dl\ based upon the metallicity cut as the dominant population is likely to extend below the cut, and conversely, there are likely to be globular clusters in the \dl\ which are more metal-rich than the metallicity cut. Hence, the metallicity cut is more likely to be indicative of where the \dl\ population of globular clusters comes to dominate over the larger galactic population. Given the best-fit model presented in Section~\ref{subsec:twocomponentmodel}, we here undertake an analysis of the posterior distributions to provide further insights into metallicity distribution of the \dl, although we readily note that this is indicative only, and more detailed modelling with more complete data is required to make robust statements. 

For each globular cluster in our sample, we assesses which population they are more likely drawn from by considering the ratio of the likelihoods, $R$, of the two models. We present this in Figure~\ref{fig:rsplit}, noting that small $R$ values correspond to globular clusters being more likely members of the more dominant galactic population rather than the \dl. For the sake of clarity in the coming analysis, we take an (arbitrary) line at $R=2$ and colour-code this sub-population of globular clusters in red. As well as a low metallicity tail, we see this sub-population has members that extend to higher metallicity.

With this separation of the populations, it is interesting to explore whether those more likely to be members of \dl\ possess clustering across any other properties. Of course, without full 3D positions and velocities, we are working within a limited representation of phase space. However, we define two quantities representing the projected angular momentum of the form;
\begin{equation}
    L_T = r v\ \ \ \ \ \ \ \ \ \ \ \ \ \ \ L_x = x v
    \label{eqn:angularmomentum}
\end{equation}
where ${x,r}$ are the tangent plane coordinates and $v$ is line-of-sight velocity as the  described in Section~\ref{subsec:data}.
These two quantities are presented in Figure~\ref{fig:pair} as a function of the metallicity, incorporating the same colour-coding with respect to $R$ as shown in Figure~\ref{fig:rsplit}. In the upper panel, it is immediately apparent that the globular clusters are more likely members of the \dl\ are found at higher total angular momentum than the dominant galactic population. This is reinforced in the lower panel of Figure~\ref{fig:pair} which presents $L_x$ as a function of metallicity, which delineates the potential members of the \dl\ from the dominant population by having large positive values of this quantity. We note that this clustering is significant in this quantity due to the alignment of the rotational axis of the \dl\ with the $y$-coordinate of the tangent plane. It is again worth emphasising that whilst this posterior analysis cannot be considered a robust method of separating the dominant and \dl\ populations, such clustering in terms of angular momentum adds credence to there being two distinct sub-populations of globular clusters present.

\begin{figure}
    \centering
    \includegraphics[width=0.95\linewidth]{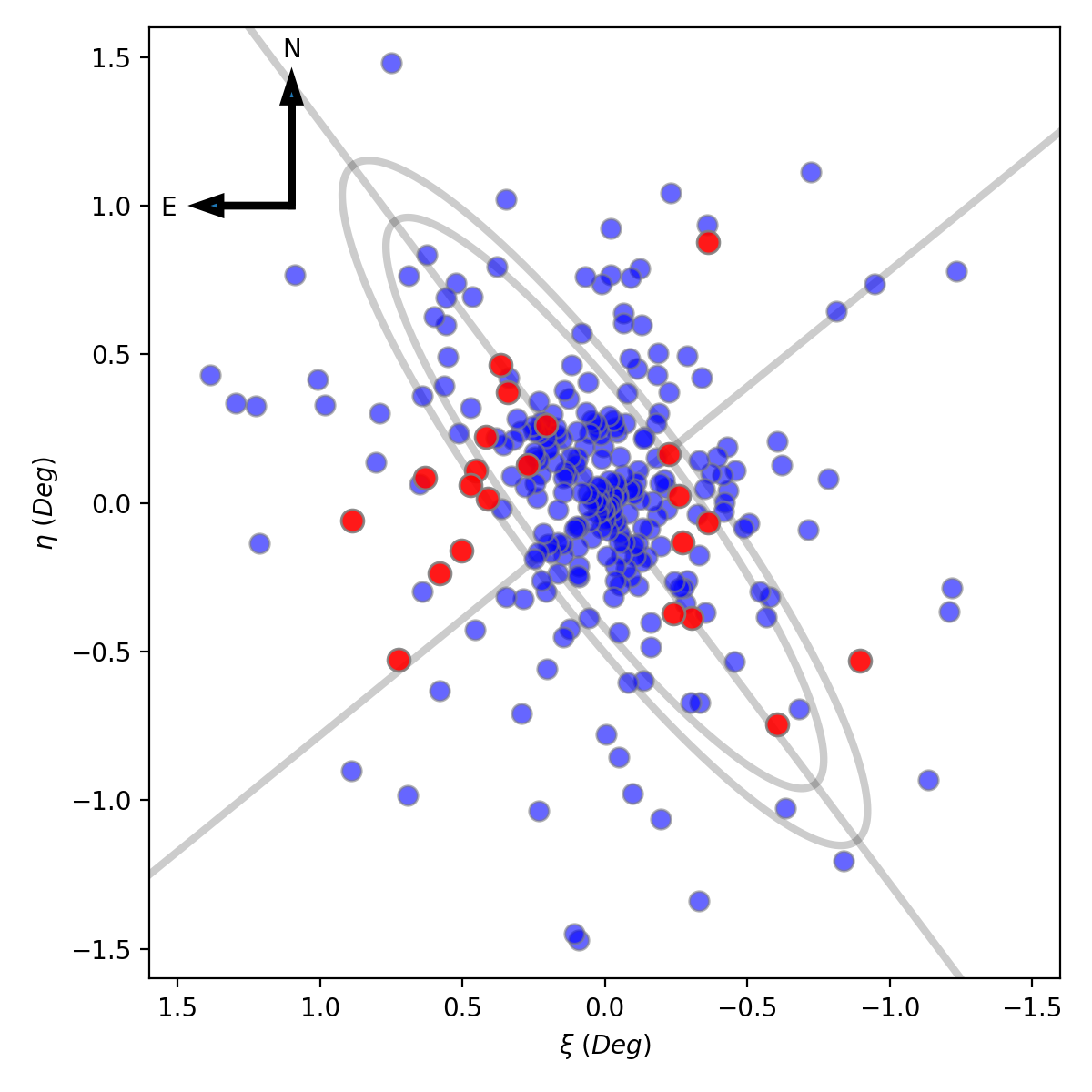}
    \caption{The spatial distribution of the globular clusters under consideration, with coordinates presented as $(\xi,\eta)$ for comparison with Figure~\ref{fig:model}. Again, the globular clusters have been coloured with respect to the relative likelihood ratio, $R$, of belonging to the galactic populations (blue) or \dl\ (red), with the division between the two sub-populations presented in Figure~\ref{fig:rsplit}.
    }
    \label{fig:position}
\end{figure}

Finally, in Figure~\ref{fig:position}, we present the spatial distribution of the globular cluster population (c.f. Figure~\ref{fig:model}) colour-coded with the relative likelihood parameter, $R$, with blue representing those most likely drawn from the dominant population, whilst those from the \dl\ are in red. Here, there is no obvious division between the two populations, and more in-depth chemical and kinematic analysis will be needed to cleanly separate the dominant population from the \dl.

\section{Conclusions}
\label{sec:conclusions}
In this paper we have presented a Bayesian evidence based analysis of the inner globular cluster population of the Andromeda galaxy, finding that two kinematic component models represent a better fit to the observed population properties than single component models. The two sub-populations are defined in relation to a metallicity cut, with the more metal-rich component being consistent with galactic rotation. However, the orbital axis of the more metal-poor component is highly offset with regards to this rotation, but is consistent with the kinematics of the outer halo globular cluster population. This inner substructure, which we name \dl, is therefore potentially part of this large-scale structure projected onto the inner part of the Andromeda Galaxy. \citet{2019Natur.574...69M} conclude that the outer halo globular clusters which are still associated with stellar substructure were accreted in the last few billion years, and hence the \dl\ represents another feature of that accretion. 
If the 10-20 globular clusters of the \dl\ do represent a single accreted object, then this would correspond to a 
progenitor halo mass of $\sim 10^{11} M_\odot$ \citep[c.f.][]{2015ApJ...806...36H}, although    
we caution that the full dynamical state and membership of this structure is essentially unknown,
\red{although the potentially large velocity dispersion found for \dl\ are suggestive of the accretion of a galaxy group into the halo of Andromeda.}

Before closing, we note that, in terms of accretion, the inner regions of Andromeda are dominated by the presence of the Giant Stellar Stream (GSS) which snakes its way around the stellar disk of the galaxy and is responsible for the stellar shells which are apparent \citep{2007ApJ...671.1591I}. Numerical simulations of the tidal formation of the GSS place the present day location of the remnants of the progenitor in front of the stellar disk of Andromeda \citep{2013MNRAS.434.2779F,2017MNRAS.464.3509K}, similar to where we find the \dl; \red{intriguingly, an estimate of the projected total angular momentum, $L_T$ (see Figure~\ref{fig:pair}) for the GSS is of order $\pm 150$ deg km/s \citep[see Figure 8 of][]{2013MNRAS.434.2779F}, similar to that seen for the \dl\, although a more detail dynamical exploration is needed to identify whether this is more than coincidence.} If the \dl\ is the remains of the progenitor of the GSS, then the angular alignment with the outer halo globular cluster population presents the intriguing possibility that the inner and outer stellar debris represents a single accretion event of a group of galaxies in the relatively recent history of the Andromeda Galaxy. This will be explored in a future contribution.

\section*{Acknowledgements}
\red{We thank Jorge Pe\~{n}arrubia and the anonymous referee for useful comments and insights that improved the quality of the paper.}
 This project was originally conceived in discussions between GFL, DM and AF, and the Bayesian exploration was planned in consultation with BJB.
 The initial kinematic investigations were undertaken as part of two separate honours projects by TA and YCL under the supervision of GFL and BJB respectively. 
 GFL has received no funding to support this research. 

\section*{Data Availability}
The data and code used in this study will be made available with a reasonable request to the authors.
The globular cluster data that formed the basis of the analysis was obtained from MMT spectroscopy, published by \citet{caldwell:16}, and  are available at \url{https://www.cfa.harvard.edu/oir/eg/m31clusters/M31_Hectospec.html}.



\bibliographystyle{mnras}
\bibliography{paper} 



\onecolumn

\appendix

\section{Catalogue of Globular Clusters}

\begin{longtable}{lccccrrccr}
\hline
Name & RA & Dec & $\xi\ (\degr)$ & $\eta\ (\degr)$ & $v$\ (km/s) & $\sigma_{v}$ & $\widehat{FeH_i}$ & $\sigma_{FeH}$ & $R$ \\
\hline
B298-G021 &  9.50092 & 40.73217 & -0.8970 & -0.5311 & -244.1 & 17.6 & -1.8 & 0.3 &  10.8 \\
B311-G033 &  9.89050 & 40.52075 & -0.6037 & -0.7459 & -222.1 & 20.3 & -1.9 & 0.2 &   5.3 \\
B336-G067 & 10.19833 & 42.14533 & -0.3606 &  0.8772 & -306.0 & 27.4 & -2.5 & 0.6 &  72.0 \\
B017-G070 & 10.20304 & 41.20197 & -0.3623 & -0.0663 & -233.9 &  3.0 & -0.8 & 0.2 &   2.8 \\
B248 & 10.28308 & 40.88361 & -0.3036 & -0.3850 & -275.3 &  3.1 & -1.4 & 0.2 &   3.0 \\
B020D-G089 & 10.32179 & 41.13586 & -0.2732 & -0.1328 & -257.1 &  3.1 & -2.1 & 0.3 &   2.5 \\
B032-G093 & 10.33962 & 41.29172 & -0.2592 &  0.0230 & -224.7 &  7.2 & -0.7 & 0.2 &   2.2 \\
B034-G096 & 10.36717 & 40.89711 & -0.2399 & -0.3717 & -255.5 & 10.2 & -0.6 & 0.2 &   2.1 \\
B036 & 10.38679 & 41.43475 & -0.2233 &  0.1659 & -208.6 &  3.0 & -0.8 & 0.2 &   2.1 \\
B198-G249 & 10.95879 & 41.53128 &  0.2053 &  0.2623 &  298.4 &  8.3 & -1.0 & 0.2 &   2.4 \\
B217-G269 & 11.04417 & 41.39756 &  0.2697 &  0.1288 &  284.6 &  3.0 & -0.8 & 0.2 &   3.0 \\
B229-G282 & 11.14096 & 41.64125 &  0.3411 &  0.3729 &  280.8 &  4.5 & -2.1 & 0.3 &   3.6 \\
B233-G287 & 11.17550 & 41.73183 &  0.3664 &  0.4636 &  228.6 & 11.6 & -1.1 & 0.2 &   2.7 \\
B283-G296 & 11.23071 & 41.28339 &  0.4104 &  0.0154 &  211.0 &  3.0 & -0.8 & 0.2 &   2.7 \\
B235-G297 & 11.24137 & 41.49000 &  0.4171 &  0.2221 &  207.8 &  3.0 & -0.9 & 0.2 &   2.6 \\
B237-G299 & 11.28842 & 41.37628 &  0.4531 &  0.1086 &  199.2 &  3.1 & -1.8 & 0.2 &   2.5 \\
B238-G301 & 11.31113 & 41.32697 &  0.4705 &  0.0594 &  257.9 &  3.0 & -0.7 & 0.2 &   4.8 \\
B240-G302 & 11.35433 & 41.10614 &  0.5047 & -0.1612 &  243.6 &  3.6 & -1.5 & 0.2 &   5.4 \\
B270D & 11.45504 & 41.03033 &  0.5812 & -0.2364 &  338.9 & 17.8 & -2.1 & 0.4 &  28.8 \\
B381-G315 & 11.52725 & 41.34967 &  0.6326 &  0.0835 &  217.5 &  4.4 & -1.1 & 0.2 &   4.1 \\
B387-G323 & 11.63963 & 40.73706 &  0.7237 & -0.5283 &  310.4 & 14.3 & -2.2 & 0.2 &  83.4 \\
B397-G336 & 11.86346 & 41.20289 &  0.8870 & -0.0604 &  179.8 & 14.7 & -1.2 & 0.2 &   2.3 \\
\hline
\caption{The 22 globular clusters with $R>2$, identified as potential members of the \dl\ (Section~\ref{sub:posterior}). For completeness, 
the metallicities and velocities for the entire sample of globular clusters are presented in Figure~\ref{fig:comparison} with the associated
error bars, demonstrating that there are no systematic issues in these quantities for those identified as being part of the \dl.
Note that the tangent plane coordinates, $(\xi,\eta)$, correspond to the $(x,y)$ coordinates used in the paper. 
}
\label{tab:table1}
\end{longtable}


\begin{figure}
    \centering
    \includegraphics[width=0.95\linewidth]{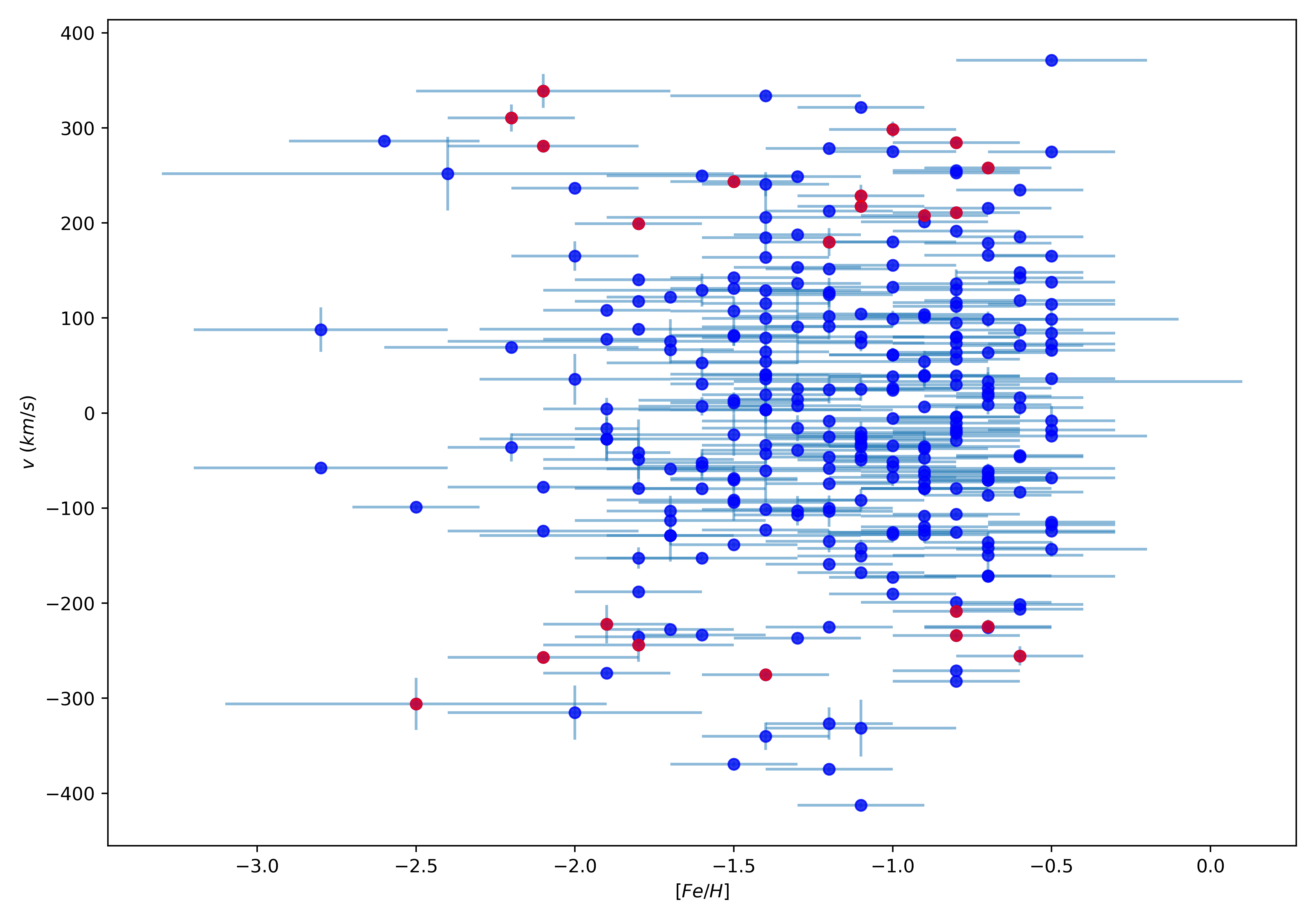}
    \caption{The globular cluster metallicities and velocities, with associated error bars, used in this study. Those with $R>2$, and hence identified as being
    most likely a member of the \dl\ in a posterior analysis (Section~\ref{sub:posterior}).
    It should be noted that there are no significant systematics with regards to the error bars for those globular clusters identified as part
    of the \dl.
    }
    \label{fig:comparison}
\end{figure}



\bsp	
\label{lastpage}
\end{document}